 \documentclass[doublespacing]{elsart}

\usepackage{amssymb}

\usepackage{cite}

\begin{document}

\begin{frontmatter}

\title{Analysis of atomic depth profiles directly extracted from Rutherford backscattering data 
for co-sputtered and ion irradiated Au-Ni films}

\author{D. Datta\corauthref{cor} \thanksref{now}} and
\corauth[cor]{Corresponding author.}
\ead{debasish.datta@saha.ac.in}
\thanks[now]{Present address: Department of Physics, Electronics and Computer Science, Seth Anandram Jaipuria College, 10, Raja Naba Krishna Street, Kolkata - 700 005, India.}
\author{S.R. Bhattacharyya}

\address{Surface Physics Division, Saha Institute of
Nuclear Physics, 1/AF Bidhan Nagar, Kolkata 700 064, India.}

\begin{abstract}
Co-sputtered Au-Ni thin films having thickness of 30 nm were deposited
on Si(100) substrates and  irradiated with 160 keV $^{40}$Ar$^{+}$ under
ambient condition at a number of fluences and analyzed using Rutherford
backscattering spectrometry (RBS). The variation of Au signal counts
in the RBS spectra with ion dose has been investigated. The distribution
of Au, Ni and Si atoms over various depths within the as deposited
and irradiated samples have been computed using the backscattering
data by means of a direct analytical method. Au and Si
profiles have been fitted with error function and the relative changes
in variance for various ion fluences compared to that of as deposited
profiles have been studied. The spreading rates of different constituents
across the interface due to Ar ion impact have also been discussed.
\end{abstract}

\end{frontmatter}

\section{Introduction}

As the dimensions of solid state devices are scaled down, it is becoming
increasingly necessary to improve the performance of the interconnects
and their electrical characteristics. Metal silicides have some advantages
over pure metal contacts due to their low-electrical resistivity and
good thermal stability and these silicides are extensively used in
microelectronic devices as contact materials to the source/drain areas
to control the Schottky barrier height \cite{Lavoie,Jiang,Wang}. Energetic
ion beams, when penetrate through the interface of different materials,
produce massive atomic transport across the interface which results
in many stable, unstable or even thermodynamically non-equilibrium
phase formation around the interface. Due to improved electrical and
chemical properties of the ion irradiated materials, mixing of metal-metal
or metal-semiconductor systems using ion beam is frequently used to
tailor different suitable contact materials for electronic devices \cite{Bibic,Liu}.
Ion bombardment induced composite formation in Au-Ni bilayer or multilayer
films deposited on Si \cite{Tamisier,Mangelinck} or III-V nitride
semiconductors\cite{Hu} have been proved to be an effective method
to fabricate contacts having low resistivity (typically, $\sim$ 
6 $\times$ 10$^{-4}$ $\Omega$-cm \cite{Kim}). 

Rutherford backscattering (RBS) technique is an easy and efficient
tool to study the dynamics and atomic rearrangement around the interface.
Analysis of RBS data and extraction of elemental depth profiles are
mostly done using some standard codes \cite{Rauhala,Doolittle,Saarilahti}
which are based on matching of experimental spectra with the theoretically
simulated spectra of hypothetical sample structures having multiple layers 
of different compositions. This type of indirect depth profiling is a difficult 
and time-consuming job sometimes leads to ambiguous results. Even for RBS 
spectra showing well separated peaks of constituent elements, this type of indirect
method to synthesize atomic depth profiles is not always a trivial issue specially 
when the number of layers in the hypothetical sample is large or when one investigates
the accumulation/depletion of a particular element over a small region inside the sample.

The aim of the present study is to implement a direct method to extract
the depth profiles of the constituents from the Rutherford backscattering
spectra and to get insight of the interactions between thin co-sputtered
Au-Ni film and Si substrate during ion bombardment. The extracted
concentration profiles of such a system and its analysis allows investigation
of the mass transport of material across the interface during Ar ion
irradiation, which may be important in the field of interconnects
and contacts in the electronic device fabrication. Specially, relatively
lower concentration of Ni in the system introduces control over the
interface dynamics and formation of Au(Ni)Si alloy generated due to
ion irradiation. In this study, we have shown the nature of impurity
spreading into the matrix and subsequent broadening of interface due
to 160 keV Ar$^{+}$ ion impact. 

\section{Experimental details}

Thin metallic films of co-sputtered Au and Ni were grown on cleaned
and polished Si(100) substrates in Ar atmosphere by dc magnetron sputtering.
Prior to the deposition, the Si substrates were etched using HF solution
to remove the native oxide layer from the surface. Before Ar introduction 
in the deposition chamber to create plasma necessary for sputtering the 
base pressure of the chamber was 8 $\times$ 10$^{-7}$ mbar. The deposition
was carried out at room temperature and the pressure of the chamber
during deposition was 9 $\times$ 10$^{-3}$ mbar. During deposition,
both the Au and Ni sources were operated simultaneously with same
power to get uniformly mixed single layer of Au and Ni over the substrates.
However, different sputtering rates of these elements gave rise to
unequal stoichiometric composition within the films with 90 at\% Au
and 10 at\% Ni. Thickness of the deposited layer was $\sim$ 30 nm
as measured using x-ray reflectivity technique. The deposited samples
were then irradiated with 160 keV $^{40}$Ar$^{+}$
ions using high current ion implanter (Danfysik) facility of Saha
Institute of Nuclear Physics.
A schematic diagram and details of the facility can be found elsewhere \cite{ISOSIIM}.
The fluences of the implanted ions were varied from 5 $\times$ 10$^{14}$
to 5 $\times$ 10$^{16}$ ions/cm$^{2}$ and the pressure
in the target chamber was 3 $\times$ 10$^{-7}$ mbar. Homogeneous
irradiation was achieved by means of a magnetic \emph{X-Y} beam scanning
system. No special arrangement was made to control the temperature
of the samples during implantation but the beam current was kept low
to avoid beam heating induced atomic transport within the sample.
Depending on the attainable ion fluences, the ion beam fluxes were
kept around 400 nA/cm$^{2}$. The projected range of Ar ions
in the target was determined using SRIM2006 code. The irradiated samples
were analyzed by means of Rutherford backscattering spectrometry (RBS)
using 2.05 MeV He$^{2+}$ ions obtained from the Pelletron
of Institute of Physics, Bhubaneswar. The backscattered data was collected
using a silicon surface barrier detector with 16 keV resolution placed
at 160$^{\circ}$. Corresponding depth profiles were extracted from
the RBS spectra by direct method described in the following section.

\section{Scheme of data analysis}

The depth profiles of the constituent elements have been extracted
directly from the backscattering data by a computer program written
using MATLAB package. The method is particularly suitable for this system 
as the peaks corresponding to different elements
are separated and well resolved. The direct determination of stoichiometry
of the ion irradiation induced atomically mixed layer ensures accurate
determination of elemental composition at various depths compared
to other indirect simulation based standard codes. However, 
using both the above-mentioned procedures for a given analysis could be very
convenient and helpful to avoid misinterpretation of the results.

Under this data analysis scheme, different elemental edges for the
incident He$^{2+}$ ion energy $E_{0}$, which is 2.05 MeV
in this case, are calculated and tracked in the experimental spectra
using the relation $k_{x}E_{0}$, where $k_{x}$'s are the kinematic
factors with $x$ representing the major elements present in the sample,
namely, Au, Ni and Si. The signal height ($H_{x}(k_{x}E_{0})$) of
an element $x$ (= Au, Ni or Si) at its respective edge is given by \cite{Chu}
\begin{equation}
H_{x}(k_{x}E_{0})=\sigma_{x}(k_{x}E_{0})Q\Omega N_{x}^{mix}\tau_{x}/\cos\theta \label{eq:signal height at surface}
\end{equation}
where $\sigma_{x}(k_{x}E_{0})$ is the differential Rutherford scattering
cross sections of element $x$ at energy $k_{x}E_{0}$, $Q$ is total
amount of charge accumulated on the sample due to He$^{2+}$
incidence during the backscattering experiment, $\Omega$ is the solid
angle subtends by the solid-state detector at the sample and $\theta$
is the angle subtends by the incident He beam with the surface normal
(in the current setup $\theta$ \textbf{}= 0$^\circ$). $N_{x}^{mix}$
is the atomic density of element $x$ in the mixture of Au, Ni and
Si at the surface and $\tau_{x}$ is the thickness of the imaginary
surface layer inside the sample which is so chosen that the energies
of He particles backscattered from this layer after colliding with
atoms of element $x$ fall within the energy window $k_{x}E_{0}-\xi$,
where $\xi$ is the energy width of one channel of the detector. It
is evident from the above fact that the depths $\tau$ for different
constituents will be different depending on the concentration of a
particular element in that mixture, respective kinematic factor and
scattering cross section. The atomic density of the mixture $N^{mix}$
is given by the relation
\begin{equation}
N^{mix}=mN_{Au}+nN_{Ni}+pN_{Si} \label{eq:N_mix}
\end{equation}
where $N$'s are the atomic densities of elements indicated by the
subscripts and \emph{m}, \emph{n} and \emph{p} are relative fractions
of Au, Ni and Si respectively. If the presence of the implanted ions
is neglected, the sum of all the atomic fraction is unity (\emph{i.e.}
\emph{m}+\emph{n}+\emph{p} = 1) and in terms of relative fraction
and atomic density, $N_{x}^{mix}$ can be expressed as
\begin{eqnarray}
&N_{Au}^{mix}&=mN_{Au}\nonumber\\
&N_{Ni}^{mix}&=nN_{Ni}\label{eq:N_mix_elements}\\
&N_{Si}^{mix}&=pN_{Si}\nonumber 
\end{eqnarray}

Using equation (\ref{eq:N_mix_elements}), the signal heights of the
different elements present on the surface (equation (\ref{eq:signal height at surface}))
can be modified as \cite{Chu}
\begin{eqnarray}
 & H_{Au}(k_{Au}E_{0}) & =\sigma_{Au}(k_{Au}E_{0})Q\Omega mN_{Au}\tau_{Au} \label{eq:Modified height Au}\\
 & H_{Ni}(k_{Ni}E_{0}) & =\sigma_{Ni}(k_{Ni}E_{0})Q\Omega nN_{Ni}\tau_{Ni} \label{eq:modified height Ni}\\
 & H_{Si}(k_{Si}E_{0}) & =\sigma_{Si}(k_{Si}E_{0})Q\Omega pN_{Si}\tau_{Si} \label{eq:Modified height Si}
\end{eqnarray}

The thicknesses corresponding to different elements expressed by $\tau_{x}$
are related to the stopping cross section factors of respective elements
$(\left[\varepsilon(E_{0})\right]_{x}^{mix})$ by the expression\cite{Khalfaoui}
\begin{equation}
\tau_{x}=\frac{\xi}{\left[\varepsilon(E_{0})\right]_{x}^{mix}N^{mix}}\label{eq:elemental tau}\end{equation}
and the weighted sum of $\tau$'s gives the effective thickness of
the layer:\begin{equation}
\tau=m\tau_{Au}+n\tau_{Ni}+p\tau_{Si}\label{eq:tau}
\end{equation}

Therefore, equations (\ref{eq:Modified height Au})-(\ref{eq:Modified height Si})
can be rewritten in terms of thickness and stopping cross section
factor as
\begin{equation}
H_{Au}(k_{Au}E_{0})=\sigma_{Au}(k_{Au}E_{0})Q\Omega mN_{Au}
\frac{\xi}{\left[\varepsilon(E_{0})\right]_{Au}^{mix}N^{mix}}\label{eq:height final}
\end{equation}
and so on.

Using the above mentioned expressions of signal heights of elements
present within the surface layer $\tau$, the ratio of relative fractions
can be expressed as\begin{equation}
\frac{m}{n}=\frac{H_{Au}(k_{Au}E_{0})}{H_{Ni}(k_{Ni}E_{0})}\frac{\sigma_{Ni}(E_{0})}{\sigma_{Au}(E_{0})}\frac{N_{Ni}}{N_{Au}}\frac{\left[\varepsilon(E_{0})\right]_{Au}^{mix}}{\left[\varepsilon(E_{0})\right]_{Ni}^{mix}}\label{eq:m/n}\end{equation}
and \begin{equation}
\frac{m}{p}=\frac{H_{Au}(k_{Au}E_{0})}{H_{Si}(k_{Si}E_{0})}\frac{\sigma_{Si}(E_{0})}{\sigma_{Au}(E_{0})}\frac{N_{Si}}{N_{Au}}\frac{\left[\varepsilon(E_{0})\right]_{Au}^{mix}}{\left[\varepsilon(E_{0})\right]_{Si}^{mix}}\label{eq:m/p}\end{equation}

The ratios $\left[\varepsilon(E_{0})\right]_{Au}^{mix}/\left[\varepsilon(E_{0})\right]_{Ni}^{mix}$
and $\left[\varepsilon(E_{0})\right]_{Au}^{mix}/\left[\varepsilon(E_{0})\right]_{Si}^{mix}$
can be treated as unity in the first approximation and using the tabulated
values of $\sigma$'s\cite{Chu}, relative concentration of each element
has been estimated.

As the concentrations of all the elements within the depth $\tau$ 
can be calculated using the above approximation, the values of $\left[\varepsilon(E_{0})\right]_{Au}^{mix}/\left[\varepsilon(E_{0})\right]_{Ni}^{mix}$
and $\left[\varepsilon(E_{0})\right]_{Au}^{mix}/\left[\varepsilon(E_{0})\right]_{Si}^{mix}$ can
be extracted using the expression of stopping cross section factor
of element $x$:
\begin{equation}
\left[\varepsilon(E_{0})\right]_{x}^{mix}=\left[\frac{k_{x}}{\cos\theta}\varepsilon^{mix}(E_{0})+\frac{1}{\cos\phi}\varepsilon^{mix}(k_{x}E_{0})\right]\label{eq:epsilon_mix_x}
\end{equation}
where $\phi$ (= $20^{\circ}$, in this experimental setup) is the
complement of scattering angle and $\varepsilon^{mix}(E_{0})$ and
$\varepsilon^{mix}(k_{x}E_{0})$ are the stopping cross sections of
the mixture at energies $E_{0}$ and $k_{x}E_{0}$ respectively. The
expressions of stopping cross section of the mixture for any given
energy $E$ is represented by 
\begin{eqnarray}
\varepsilon^{mix}(E) & = & m\varepsilon_{Au}(E)+n\varepsilon_{Ni}(E)+p\varepsilon_{Si}(E)\label{eq:epsilon_mix}
\end{eqnarray}
where $\varepsilon_{Au}(E)$, $\varepsilon_{Ni}(E)$ and $\varepsilon_{Si}(E)$
are individual stopping cross sections of Au, Ni and Si respectively
which are evaluated by the polynomial fit expressions \cite{Chu}
\begin{equation}
\varepsilon(E)=\sum_{i=0}^{5}A_{i}E^{i}\label{eq:epsilon}
\end{equation}
with $A_{0}-A_{5}$ being constants for a particular element when
$E$ is expressed in MeV.

Using equations (\ref{eq:epsilon_mix_x}) and (\ref{eq:epsilon_mix}),
the stopping cross section factors for different elements are calculated
and fed in equations (\ref{eq:m/n}) and (\ref{eq:m/p}) to calculate
exact values of atomic fractions which values are again used in equations
(\ref{eq:epsilon_mix_x}) and (\ref{eq:epsilon_mix}) in iterative
manner to produce more accurate values of elemental concentrations
in the layer.

As the thickness $(\tau)$ and composition $(N^{mix})$ of the first
layer is known, the energy, loses by the penetrating He ion while
traveling through that layer can also be estimated and the energy
of the ion just before entering the next layer $(E)$ is given by
\begin{equation}
E=E'-\frac{\varepsilon^{mix}(E')N^{mix}\tau}{\cos\theta}\label{eq:energy at next layer}
\end{equation}
where $E'$ represents the energy of the ion at the point of entering
into the previous layer (= $E_{0}$, when dealing with second layer
starting from surface). Simultaneously, signal heights of each element
in the next lower channels starting from the corresponding elemental
edges are also extracted. The ratios of atomic fractions in the second
layer are given by
\begin{eqnarray}
 \frac{m}{n}=\frac{H_{Au}(E_{Au})}{H_{Ni}(E_{Ni})}\frac{\sigma_{Ni}(E)}{\sigma_{Au}(E)}\frac{N_{Ni}}{N_{Au}} \left\{ \frac{\left[\varepsilon(E)\right]_{Au}^{mix}}{\left[\varepsilon(E)\right]_{Ni}^{mix}}\frac{\varepsilon(E_{Ni})}
{\varepsilon(E_{Au})}\frac{\varepsilon(k_{Ni}E)}{\varepsilon(k_{Au}E)}\right\} \label{eq:m/n depth}
\end{eqnarray}
and
\begin{eqnarray}
 \frac{m}{p}=\frac{H_{Au}(E_{Au})}{H_{Si}(E_{Si})}\frac{\sigma_{Si}(E)}{\sigma_{Au}(E)}\frac{N_{Si}}{N_{Au}}\left\{ \frac{\left[\varepsilon(E)\right]_{Au}^{mix}}{\left[\varepsilon(E)\right]_{Si}^{mix}}\frac{\varepsilon(E_{Si})}
{\varepsilon(E_{Au})}\frac{\varepsilon(k_{Si}E)}{\varepsilon(k_{Au}E)}\right\} \label{eq:m/p depth}
\end{eqnarray}
where $H_{Au}(E_{Au})$ is the height of Au signal in the channel
preceding to that which corresponds to Au edge and $E_{Au}$ is the
energy equivalent of that particular channel. Similarly, $H_{Ni}(E_{Ni})$
and $H_{Si}(E_{Si})$ is the heights for Ni and Si signals respectively
at channels next to Ni and Si edges towards the lower energy side.
$\varepsilon(E_{Au})$ and $\varepsilon(k_{Au}E)$ \emph{etc.} are
stopping cross sections at energies $E_{Au}$ and $k_{Au}E_{Au}$
and so on. $\left[\varepsilon(E)\right]_{x}^{mix}$ is the stopping
cross section factor for element $x$ present in the mixture and expressed
by the same relation as equation (\ref{eq:epsilon_mix_x}) with energy
$E_{0}$ replaced by $E$.

When treating the second layer, the approximation applied to the surface
layer for determining the elemental composition is also valid which
treats the values of the quantities within curly braces in equations
(\ref{eq:m/n depth}) and (\ref{eq:m/p depth}) as unity. So, first
estimation of the concentrations can be figured out using that approximation
which, in turn, can be used along with equation (\ref{eq:epsilon})
and tabulated values of $\sigma$'s\cite{Chu} in iterative manner
to determine accurate values of elemental concentrations in that layer.
Subsequent determination of $N^{mix}$ and $\tau$ for the layer can
be done using equations (\ref{eq:N_mix}), (\ref{eq:elemental tau})
and (\ref{eq:tau}). For estimating elemental concentrations of third,
forth and subsequent deeper layers, same formalism and expressions
as applied to second layer can be followed.

It should be noted that experimentally obtained RBS spectra are convoluted
with the apparatus function which contains initial beam energy dispersion, 
energy resolution of the detector, energy loss and energy-loss straggling 
in the detector as well as in the sample. But the determination of deconvoluted
spectra using the normal deconvolution process is not trivial and sometimes introduces
unwanted undulations in the spectra which may yield negative concentration of an 
element at some depth. So some other approaches described in ref.\cite{Ellmer} or
Bayesian probability theory \cite{Fischer} are suitable for recovering the 'true' 
spectra. Since the final aim of this paper is to investigate the relative broadening 
of Au and Si profiles with increased ion fluence in case of co-sputtered Au-Ni system, 
the effect of apparatus function causes unidirectional change in the RBS spectra 
and hence nullified.

\section{Results and discussion}

Figure 1 represents Rutherford backscattering
spectra of as deposited and samples irradiated
with Ar ions at different fluences. It can be observed from the figure
that the Au signal height decreases with increasing dose which signifies
sputter erosion of the films. On the other hand, Ni signals do not
decrease significantly with increasing dose. This is due to the fact
that higher sputtering yield of gold triggers preferential sputtering
of Au atoms within the mixture of Au and Ni. As a result, Ni signal
heights decrease less rapidly than that of Au. Along with diminution
of Au and Ni peaks, tailing of these signals towards lower energy
side can also be observed representing ion irradiation induced atomic
diffusion of constitutes in the direction of sample interior. The
accumulation of Ar atoms which penetrate into the samples during ion
irradiation, show no considerable change in the spectra until a dose
of 1 $\times$ 10$^{16}$ ions/cm$^{2}$ beyond which
a strong peak of Ar emerged between Ni and Si peaks. In contrast to
the spreading of Au and Ni signals towards Si side, Ar peak at a fluence
of 5 $\times$ 10$^{16}$ ions/cm$^{2}$ shows symmetry
around its maximum position implying Gaussian type distribution commonly
observed in case of implanted ions in solids at relatively low doses.

Variation of total counts enveloped by the Au signals in the RBS spectra
with ion dose have been showed in figure 2. The
total count for as deposited sample has been taken as unity and other
counts are normalized with respect to that. The counts decrease rapidly
at the initial stage of ion irradiation and after achieving a dose
of 1 $\times$ 10$^{16}$ ions/cm$^{2}$ it becomes
monotonous which is in accordance with the previously reported result
regarding sputter erosion of thin films\cite{Au}. The data has been
fitted using exponential decay function\begin{equation}
C=C_{0}+A\exp\left(-\frac{\textrm{fluence}}{B}\right)\label{eq:exponential decay}\end{equation}
where $C$ and $C_{0}$ are the normalized and saturated counts respectively
and $A$, $B$ are constants. The fit gives the values of different
parameters as $C_{0}=$ 0.62 $\pm$ 0.03, $A=$ 0.36 $\pm$ 0.04 and
$B=$ (27.98 $\pm$ 11.09)  $\times$ 10$^{14}$ ions/cm$^{2}$.
The exponential decay of the Au counts with ion dose suggests that
sputtering of thin films is different from that of bulk material.
At the early phase of ion bombardment the sputtering is to some extent
analogous to bulk sputtering and the sputtering yields of Au and Ni
for an ion-target combination as the present one are $\sim$ 4.0 and
$\sim$ 0.5 atoms/ion respectively, as calculated using SRIM code.
The outcome of larger yield for Au is a Ni-rich top layer along with
surfacing of Si due to ion induced cascade mixing of the components
which produces a saturated region in figure 2\ref{fig:Au counts}  above
a critical dose of 1 $\times$ 10$^{16}$ ions/cm$^{2}$.

The depth profiles of as deposited and irradiated samples as shown
in figure 3 reveal the possible cause of exponential
type decaying of integrated Au counts represented in figure 2.
The depth profiles of as deposited and sample irradiated at a dose
of 5 $\times$ 10$^{14}$ ions/cm$^{2}$ show no significant
difference with a presence of $\sim$ 90\% Au and $\sim$ 10\% Ni
at the surface. After achieving a fluence of 1 $\times$ 10$^{15}$
ions/cm$^{2}$, the sputter erosion of top surface layer is
visible in the Au profile which is more noticeable at the next dose
by the advancement of Si profile towards the surface. The Si concentration
at the surface has reached around 15\% for ion dose of 
5 $\times$ 10$^{15}$ ions/cm$^{2}$, with a reduction of the amount
of Au. At higher doses the effect of preceding of Si profile is more
dominant. Moreover, at a fluence of 1 $\times$ 10$^{16}$ ions/cm$^{2}$,
a surface layer of 10 nm having an approximate composition of Au$_{58}$Ni$_{5}$Si$_{37}$
covering another layer of Au$_{45}$Ni$_{5}$Si$_{50}$
which spreads over $\sim$ 20 nm has emerged. At the next higher ion
dose, \emph{i.e}. at 5 $\times$ 10$^{16}$ ions/cm$^{2}$,
the buried compound layer has been exposed leaving completely mixed
surface layer having a thickness of $\sim$ 20 nm. The formation of
mixed layers at the surface and subsurface regions at higher irradiation
doses are responsible for saturation of Au signal count represented
in figure 2. However, the Ni profiles at various
fluences show very little observable changes due to irradiation.

The extracted atomic profiles has been used to quantify the growth
of the mixed region due to ion irradiation. Au and Si profiles are
fitted with error function to deduce the change in variance $\sigma^{2}$
as a function of ion dose. The relative change in variances ($\Delta\sigma^{2}$)
for various doses with respect to the variance of Au profile for as
deposited sample is represented in figure 4. A linear
growth of the variance with the ion fluence had been observed previously
in many metal/semiconductor systems due to ion irradiation\cite{Was,Milinovic,Gold,Nastasi}.
But for the present system, the linear variation of $\Delta\sigma^{2}$
with ion dose seems to be valid only up to a fluence of 
1 $\times$ 10$^{16}$ ions/cm$^{2}$. Beyond that $\Delta\sigma^{2}$
for both Au and Si profiles decrease to a lower value. The linear
fit of present data up to 1 $\times$ 10$^{16}$ ions/cm$^{2}$
give the mixing rates $\Delta\sigma^{2}/\textrm{fluence}$ = 6.88
$\pm$ 0.77 nm$^{4}$ for Au and 9.34 $\pm$ 0.43 nm$^{4}$ for Si.
The difference in mixing rates is probably due to the diffusion of
Ni atoms in Si layer which occurs simultaneously with Au. The lower
heat of compound formation (--16 kJ/g at.) for Ni and Si and hence
ready creation of different Ni-Si phases forms a physical barrier
before the Au atoms to spread at an equal rate of Si. The decrease
in $\Delta\sigma^{2}$ beyond 1 $\times$ 10$^{16}$ ions/cm$^{2}$
can be explained with the help of figure 2 and 3. The SRIM simulated range of impinging Ar
ions is $\sim$ 125 $\pm$ 65 nm which is sufficient to deposit considerable
energy on both sides of the interface and produces collision cascades.
Collision cascades both in film and substrate regions set off recoil
implantation, cascade mixing and radiation enhanced diffusion processes.
Simultaneous preferential surface sputtering and cascade mixing not
only form a Au-depleted surface but also an accumulation of Si around
the near-surface region at and beyond 1 $\times$ 10$^{16}$ ions/cm$^{2}$
Ar dose (figures 3(e) and 3(f)).
Eventually, the evolution of homogeneously mixed layer of Au$_{45}$Ni$_{5}$Si$_{50}$
which has been already discussed, contributes towards the sputtering
process. The notable fact about these mixed layers is the absence
of preferential sputtering of any species due to ion impact. The major
components of the mixture, namely Au and Si are sputtered at the same
rates which is supported by the SRIM simulation of similar kind of
target composition-ion combination. The simulation provides the erosion
yields of both Au and Si as $\sim$ 1.5 atoms/ion and as a result,
not any single component but the whole surface is sputtered away with
ion impact. This fact is further supported by figure 2
which shows a flat region past a threshold dose. Thus, the bulk Si
is becoming exposed which gives rise to lower values of $\Delta\sigma^{2}$
at an ion fluence of 5 $\times$ 10$^{16}$ ions/cm$^{2}$.

\section{Conclusion}

In this study, an efficient analytical method to analyze RBS data
has been presented. Using that method, depth profiles of different
components of co-sputtered and ion irradiated Au-Ni thin films on
Si has been investigated. It has been observed that variation of $\Delta\sigma^{2}$
with dose follows a linear relationship up to a certain ion fluence
which yields mixing rates of 6.88 $\pm$ 0.77 and 9.34 $\pm$ 0.43
nm$^{4}$ for Au and Si, respectively. The lowering of $\Delta\sigma^{2}$
values ahead of a critical dose can be explained by the formation
of Au-Ni-Si composite at the surface from which elements are sputtered
out evenly which is further supported by variation of Au counts in
RBS spectra with ion dose.

\section{Acknowledgement}

The authors are indebted to Mr. S. Roy for the assistance in depositing film 
using the magnetron sputtering unit. They also thank all members of Ion Beam 
Laboratory of Institute of Physics, Bhubaneswar for RBS measurements.



\newpage

{\bf Figure captions}

{\bf Figure 1:} RBS spectra of (a) as deposited co-sputtered Au-Ni sample and samples
irradiated using 160 keV $^{40}$Ar$^{+}$ with doses of (b)
5 $\times$ 10$^{14}$, (c) 1 $\times$ 10$^{15}$,
(d) 5 $\times$ 10$^{15}$, (e) 1 $\times$ 10$^{16}$
and (f) 5 $\times$ 10$^{16}$ ions/cm$^{2}$. The upper and lower panels represent the same spectra but in the lower panel each spectrum is vertically shifted from others. The spectra
are normalized with respect to Si signal height and different scales
are used for Au and Ni/Si signals for clarity.\label{fig:RBS spectra}

{\bf Figure 2:} Integrated counts under Au signals in the RBS spectra of co-sputtered
films. The counts are normalized with respect to that of as deposited
sample. The data has been fitted using the equation of exponential
decay.\label{fig:Au counts}

{\bf Figure 3:} Depth profiles of (a) as deposited co-sputtered Au-Ni sample and
samples irradiated using 160 keV $^{40}$Ar$^{+}$ with doses
of (b) 5 $\times$ 10$^{14}$, (c) 1 $\times$ 10$^{15}$,
(d) 5 $\times$ 10$^{15}$, (e) 1 $\times$ 10$^{16}$
and (f) 5 $\times$ 10$^{16}$ ions/cm$^{2}$.\label{fig:Depth profile}

{\bf Figure 4:} Change of variance of (a) Au and (b) Si profiles as a function of
ion dose and linear fit (dotted line) of the experimental data.\label{fig:variance}

\end{document}